\documentclass[aps,pra,showpacs]{revtex4}

\usepackage{graphicx}

\begin{document}

\title{Pulse shape effects on photon-photon interactions in non-linear 
optical quantum gates}
 
\author{Holger F. Hofmann}
\email{hofmann@hiroshima-u.ac.jp}
\author{Hitoshi Nishitani} 

\affiliation{Hiroshima University, 1-3-1 Kagamiyama, Higashihiroshima, Hiroshima 739-8530, Japan}

\date{\today}

\begin{abstract}
Ideally, strong non-linearities could be used to implement quantum gates for photonic qubits by well controlled two photon interactions. However, the dependence of the non-linear interaction on frequency and time makes it difficult to preserve a coherent pulse shape that could justify a single mode model for the time-frequency degree of freedom of the photons. In this paper, we analyze the problem of temporal multi-mode effects by considering the pulse shape of the average output field obtained from a coherent input pulse. It is shown that a significant part of the two photon state transformation can be derived from this semi-classical description of the optical non-linearity. The effect of a non-linear system on a two photon state can then be determined from the density matrix dynamics of the coherently driven system using input-output theory.  As an example, the resonant non-linearity of a single two level atom is characterized. The results indicate that the most efficient non-linear effect may not be the widely studied single mode phase shift, but the transfer of one of the photons to an orthogonal mode distinguished by its temporal and spectral properties.
\end{abstract}

\maketitle

\section{Introduction}

Photonic qubits based on the polarization or the transverse mode structure of single photons are an attractive candidate for the implementation of quantum information protocols because it is relatively easy to establish and maintain single qubit coherences by conventional linear optics. However, the realization of well-controlled two photon interactions remains a challenging problem on the road to larger networks and more efficient operations. 
Early on, it has been suggested that non-linear materials might provide the
interaction necessary to couple pairs of photons \cite{Imo85}, and 
sufficiently strong non-linear effects were demonstrated experimentally
using single atoms in a high-finesse cavity \cite{Tur95}. Recent advances
in solid state cavity designs seem to be putting the prospect of integrated 
devices implementing optical non-linear quantum gates within the reach of 
present technological capabilities \cite{Aok06,Cha07,Tak08,Fus08,Min09}. However, 
some quite fundamental problems still need to be addressed before the proper functions of a quantum gate can be realized. In particular, it is necessary to 
preserve single mode coherence, not only in polarization and in
transverse mode structure, but also in the time-frequency domain. The latter problem is fundamentally linked to the dynamics of optical non-linearities in
time and space \cite{Hof03a,Koj03,Hof03b,Kos04,Wak06,Auf07,Kos07,Koj07,Kos08,Xia08}, and has recently 
been identified as a critical problem in the realization of quantum gates 
\cite{Sha06,Sha07,Leu08}. 
Initial attempts at optimizing the pulse shape and the pulse durations focused on the possibility
of obtaining a large phase flip, represented by a negative two photon amplitude in a single intended
target mode \cite{Kos04,Koj04}. In that context, Koshino and Ishihara pointed out that the two photon
amplitude can be evaluated from the semi-classical response of the system \cite{Kos04}. This result 
suggests that a more detailed analysis of the relation between semi-classical field expectation 
values and the transformation of two photon wavefunctions may be possible. Such an analysis 
could provide both a more thorough foundation for the evaluation of experimental results like 
the ones recently reported in \cite{Che06,Mat09} and a more detailed characterization
and classification of the spectral and temporal features observed in highly non-linear devices.
In the following, we therefore present a systematic analysis of the relation between the average
field response obtained from a coherent input and the transformation of the two photon component
of the quantum mechanical wavefunction. 

In section \ref{sec:system}, a quantum mechanical formulation
of the semi-classical non-linear response is developed using field operators. In section 
\ref{sec:quantum}, the non-linear interaction is expanded in terms of its effects on photon number
states of suitably defined modes. It is then possible to express the semi-classical output in terms
of the wavefunctions of these modes. In section \ref{sec:derivation}, it is shown how the matrix
elements of the photon number expansion can be derived from the overlap integrals of the
semi-classical result.  According to the results of sections \ref{sec:system} to \ref{sec:derivation},
the non-linear transformation has two distinct effects: a change in phase and amplitude
of the two photon component of the linear output pulse and a change in pulse shape represented by the 
transfer of a single photon to an orthogonal mode. In section \ref{sec:tdependence}, the magnitude of these
effects is investigated for the case of a resonant two-level system. It is shown that  
the conditional transfer of one photon to an orthogonal mode is much stronger than the
non-linear phase shift, suggesting that it may be more efficient to use this effect in optical
quantum gates. In section \ref{sec:pulse}, the pulse shapes of the modes involved in the non-linear
photon tranfer are presented and the dynamics of the effect is discussed. The conclusions 
are summarized in section \ref{sec:conclusions}.

\section{Semi-classical characterization of a non-linear quantum system}
\label{sec:system}

In order to analyze the quantum level effects of an optical non-linearity,
we start by considering the effects of a quantum system on the pulse 
shape of a coherent input pulse. Such an input pulse can be described by
a classical time dependent field amplitude $\alpha \; b_{\mbox{\small in}}(t)$,
where $b_{\mbox{\small in}}(t)$ is the wave function of a single mode 
light field normalized to one photon. Hence, the input quantum 
state is effectively a single mode coherent state $\mid \! {\alpha}\rangle$. 
This state interacts with the non-linear quantum system, temporarily 
exciting it from its initial ground state. After this interaction, the 
system returns to the ground state, re-emitting any photons it might have
absorbed in the process of the interaction. 

\begin{figure}[th]
\begin{picture}(400,150)
\put(0,0){\makebox(400,150)
{\vspace*{-5cm}
\scalebox{0.8}[0.8]{\includegraphics{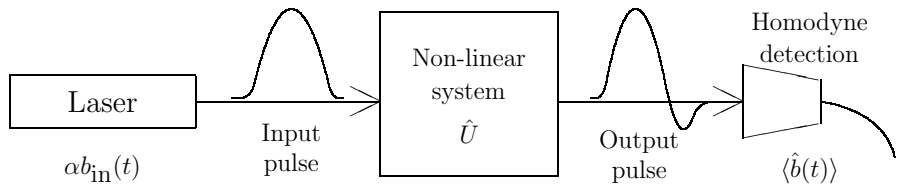}}}}
\end{picture}
\caption{\label{fig1}
Illustration of the semi-classical characterization of an optically
non-linear system using coherent input light and homodyne detection
of the output pulse shape.}
\end{figure}

If decoherence effects can be
neglected, the total effect of the interaction on the quantum state of the
input pulse can then be written
as a unitary transformation $\hat{U}$ acting only on the state of the 
light field. In general, this unitary transformation acts on the full 
continuum of modes in free space. 
A complete characterization of the output
state would therefore require multi-photon coherences between all 
frequencies or times observed in the output \cite{Koj03,Kos08,Sha06}. 
However, we now simplify this analysis by considering only the expectation
values of the output fields $\langle \hat{b}(t) \rangle$. 
Experimentally, this would correspond to the averages of a 
homodyne measurement, as used in the pioneering work of Turchette et al.
\cite{Tur95}. Fig. \ref{fig1} shows a schematic illustration of this semi-classical
characterization of a quantum level non-linearity.
Theoretically, such averages are easily obtained from the
density matrix dynamics of the system using input-output theory, as
will be explained in more detail in section \ref{sec:dynamics}. 

In terms of the unitary operation $\hat{U}$ describing the effects of the
non-linear system on the quantum state of the input light, the average 
output field is given by the expectation value of the field operator $\hat{b}(t)$ in 
the time domain,
\begin{equation}
\label{eq:avout}
b_{\mbox{\small out}}(t) = \langle \hat{b}(t) \rangle = 
\langle \alpha \mid \hat{U}^\dag\hat{b}(t)\hat{U} \mid \alpha \rangle
\end{equation}
We can then proceed to analyze the specific form of the output pulse
$b_{\mbox{\small out}}(t)$ by expanding the output in terms of the input amplitude
$\alpha$. The non-linear effect arising from photon-photon interactions is given by 
the third-order term in this expansion. For sufficiently small amplitudes $\alpha$, 
the semi-classical effect of photon-photon interactions on the transformation 
of an input pulse $\alpha \; b_{\mbox{\small in}}(t)$ can therefore be expressed
in terms of the third-order non-linear respone,
\begin{equation}
\label{eq:nlout}
b_{\mbox{\small out}}(t)=\alpha b^{(1)}(t)+\alpha|\alpha|^2 b^{(3)}(t),
\end{equation}
where $b^{(1)}(t)$ is the pulse shape of the linear output, 
and $b^{(3)}(t)$ is the pulse shape of the third-order non-linearity.

If there is neither decoherence nor photon loss in the system, 
$b^{(1)}(t)$ describes a normalized single 
output mode that characterizes the transformation of the one 
photon wavefunction from 
$\psi_{\mbox{\small in}}(t)=b_{\mbox{\small in}}(t)$ to
$\psi_1(t)=b^{(1)}(t)$. On the other hand, $b^{(3)}(t)$ is neither 
normalized nor does it represent a wavefunction orthogonal to 
$\psi_1(t)$. However, it is possible to interpret the pulse shape
$b^{(3)}(t)$ as a linear superposition of a component 
proportional to $\psi_1(t)$ and another component proportional to
a normalized wavefunction $\psi_2(t)$ that is orthogonal to 
$\psi_1(t)$. Since the wavefunctions $\psi_1(t)$ and $\psi_2(t)$ 
are orthogonal, they represent two distinct modes in the time-frequency
continuum. It is therefore possible to quantize the light field
by assigning separate annihilation operators
$\hat{a}_1$ and $\hat{a}_2$ to these orthogonal modes. 
The output pulse shape $b_{\mbox{\small out}}(t)$ can then be 
expressed in terms of the two orthonormal wavefunctions 
$\psi_1(t)$ and $\psi_2(t)$ and the expectation values $\langle \hat{a}_1 \rangle$ 
and $\langle \hat{a}_2 \rangle$ of their
complex amplitudes,
\begin{equation}
\label{eq:quantout}
b_{\mbox{\small out}}(t)=\langle \hat{a}_1 \rangle 
\psi_1(t)+\langle \hat{a}_2 \rangle \psi_2(t).
\end{equation}
The expectation values can be determined by analyzing the pulse shape
functions $b^{(1)}(t)$ and $b^{(3)}(t)$ in eq.(\ref{eq:nlout}). 
Thus, the semi-classical representation of the non-linearity in terms of
field expectation values can be interpreted within a fully quantum 
mechanical model focussing on the two modes defined by the pulse shapes 
observed in the output field.

\section{Quantum mechanics of the photon-photon interaction}
\label{sec:quantum}

In the previous section, we have shown that the third-order non-linear 
response to a coherent input pulse can be described in terms of
two quantized modes defined by the linear and non-linear parts of
the average output field $b_{\mbox{\small out}}(t)$.
We can now use this two mode representation to formulate the
matrix elements of the unitary transformation $\hat{U}$ 
that describe the transitions between the photon number states 
of these modes.
If photon losses can be neglected, the unitary transformation preserves 
the total photon number and we can look at each subspace of fixed total photon
number separately. In particular, the vacuum state will
not be changed by the interaction with the system, so the effect of 
$\hat{U}$ on the zero photon subspace is simply given by
\begin{equation}
\label{eq:vtv}
\hat{U} \mid \mbox{vac.} \rangle=\mid \mbox{vac.} \rangle.
\end{equation}
For low intensity fields ($\alpha \ll 1$), the expectation values 
$\langle \hat{b}(t) \rangle$ 
of the field amplitudes are given by the coherence 
between the single photon wavefunction at $t$ and the vacuum. Therefore, 
the linear part of the semi-classical pulse shape transformation is 
equivalent to the transformation of the single photon wavefunction.
Specifically, the single photon input wavefunction 
$\psi_{\mbox{\small in}}(t)=b_{\mbox{\small in}}(t)$ is transformed 
into the linear output mode wavefunction $\psi_1(t)=b^{(1)}(t)$. 
In terms of the annihilation operators of the input mode 
$\hat{a}_{\mbox{\small in}}$ and the linear output mode 
$\hat{a}_1$, the effect of the unitary transformation $\hat{U}$ 
in the single photon subspace can therefore be expressed as 
\begin{equation}
\label{eq:1t1}
\hat{U}\left(\hat{a}^\dag_{\mbox{\small in}} 
\mid \mbox{vac.} \rangle \right)=\hat{a}_1^\dag \mid \mbox{vac.} \rangle.
\end{equation}

The transformation in the two photon subspace is more complicated,
because the photon-photon interaction generally entangles the wavefunctions
of the two photons \cite{Hof03b}. However, we know from 
eq.(\ref{eq:nlout}) that a significant part of the two photon output 
wavefunction can be described in terms of the modes $\hat{a}_1$ and
$\hat{a}_2$. We can therefore expand the effect of $\hat{U}$ in the
two photon subspace in terms of these output modes.
Specifically, the contributions to the expectation values of the 
annihilation operators in eq.(\ref{eq:quantout}) originate from an
inner product of a one photon wavefunction generated by annihilating a
photon from the output two photon wavefunction and the output one photon 
state. As a result, the components of the two photon output
contributing to the averages in  eq.(\ref{eq:quantout}) must have at least one photon
in the linear output mode $\hat{a}_1$. The components of the two photon
output that contribute to $b_{\mbox{\small out}}(t)$ are therefore
(i) a state where both photons are in mode $\hat{a}_1$ and 
(ii) a state where one photon is in mode $\hat{a}_1$ and one photon
is in mode $\hat{a}_2$. In addition, there is a third component 
$\mid \mbox{rest} \rangle$ that does not contribute to 
$b_{\mbox{\small out}}(t)$ because it does not have any photons in the
linear output mode $\hat{a}_1$. Thus, the expansion of the two photon
output state can be written as
\begin{equation}
\label{eq:2t2}
\hat{U}\left(\frac{1}{\sqrt{2}}\hat{a}_{\mbox{\small in}}^\dag\hat{a}_{\mbox{\small in}}^\dag \mid{\mbox{vac.}}\rangle\right)
= \frac{C_{11}}{\sqrt{2}}\hat{a}_{1}^\dag\hat{a}_{1}^\dag\mid{\mbox{vac.}}\rangle
+C_{12}\hat{a}_{1}^\dag\hat{a}_{2}^\dag\mid{\mbox{vac.}}\rangle
+C_r\mid{\mbox{rest}}\rangle,
\end{equation}
where $C_{11}$, $C_{12}$, and $C_r$ are the two photon amplitudes
characterizing the photon-photon interaction described by $\hat{U}$.
$C_{11}$ describes the amplitude of obtaining both photons in the same
mode as the single photon output, $C_{12}$ describes a non-linear
transfer of one photon to a well defined orthogonal mode, and 
$C_r$ describes the amplitude of processes where both photons are 
transfered to other modes with shapes that cannot be identified using only
the semi-classical output average $b_{\mbox{\small out}}(t)$. 
 
We can now solve eq.(\ref{eq:avout}) by applying the unitary 
transformation $\hat{U}$ to the single mode coherent input state 
$\mid{\alpha}\rangle$. In order to describe the third-order non-linearity, it is
convenient to expand the coherent state, neglecting all terms that only 
contribute terms of fourth or higher order in $\alpha$ to the final field
expectation value in eq.(\ref{eq:avout}). The coherent state can then be
approximated by
\begin{equation}
\mid{\alpha}\rangle\approx\left(1-\frac{|\alpha|^2}{2} \right)\mid{\mbox{vac.}}\rangle
+\alpha \left(1-\frac{|\alpha|^2}{2} \right)\hat{a}^\dag_{\mbox{\small in}}\mid{\mbox{vac.}}\rangle
+\frac{\alpha^2}{2}\hat{a}_{\mbox{\small in}}^\dag\hat{a}_{\mbox{\small in}}^\dag 
\mid{\mbox{vac.}}\rangle.
\end{equation}
The application of $\hat{U}$ to this input state results in an output
state of
\begin{equation}
\hat{U}\mid{\alpha}\rangle
 =\left(1-\frac{|\alpha|^2}{2}\right)\mid{\mbox{vac.}}\rangle+\alpha\left(1-\frac{|\alpha|^2}{2}\right)\hat{a}^\dag_{1}\mid{\mbox{vac.}}\rangle
 +\frac{\alpha^2}{\sqrt{2}} \left(\frac{C_{11}}{\sqrt{2}}\hat{a}^\dag_{1}\hat{a}^\dag_{1}\mid{\mbox{vac.}}\rangle
 +C_{12}\hat{a}^\dag_{1}\hat{a}^\dag_{2}\mid{\mbox{vac.}}\rangle
 +C_r\mid{\mbox{rest}}\rangle\right).
\end{equation}
From this output state, we can obtain the expectation values of $\hat{a}_1$
and $\hat{a}_2$,
\begin{eqnarray}
\label{eq:ampout}
 \langle \hat{a}_1 \rangle &=& \langle \psi_{\mbox{\small out}} \mid\hat{a}_1 \mid \psi_{\mbox{\small out}} \rangle=\alpha+(C_{11}-1)\alpha|\alpha|^2 \nonumber \\
 \langle{\hat{a}_2}\rangle&=& \langle \psi_{\mbox{\small out}} \mid
\hat{a}_2 \mid \psi_{\mbox{\small out}} \rangle = \frac{C_{12}}{\sqrt{2}}\alpha|\alpha|^2.
\end{eqnarray}
These expectation values establish the connection between the few mode formulation of $b_{\mbox{\small out}}(t)$ in eq.(\ref{eq:quantout}) and the
expansion up to third-order in $\alpha$ given by eq.(\ref{eq:nlout}). 
Specifically, the expression for $b_{\mbox{\small out}}(t)$ obtained 
by inserting the results given by eq.(\ref{eq:ampout}) into eq.(\ref{eq:quantout}) reads
\begin{equation}
\label{eq:ciout}
 b_{\mbox{\small out}}(t) = \alpha\psi_1(t)+\alpha|\alpha|^2
\left( (C_{11}-1)\psi_1(t)+
\frac{C_{12}}{\sqrt{2}}\psi_2(t)\right).
\end{equation}
Based on this relation, it is possible to obtain the two photon amplitudes $C_i$ that characterize
the unitary transformation $\hat{U}$ by decomposing the non-linear output 
pulse shape $b^{(3)}(t)$ of eq.(\ref{eq:nlout}) into its $\psi_1(t)$ and $\psi_2(t)$ components.

\section{Derivation of two photon amplitudes $C_i$ from 
semi-classical pulse shapes}
\label{sec:derivation}

The two photon amplitudes $C_i$ characterize the 
essential properties of the 
non-linear system for applications as an optical quantum gate.
In particular, $C_{11}$ describes any non-linear phase shift, with 
$C_{11}=-1$ corresponding to the ideal controlled phase-flip needed
for the implementation of a quantum controlled-NOT \cite{Hof03a,Kos04,Koj04}. 
On the other hand, $C_{12}$ describes a well-controlled transfer of one
photon to a new mode. Since the output is fully quantum coherent, this
process may also be a suitable candidate for quantum information processing.
Finally, the coefficient $C_r$ represents a component of unknown 
coherence that may be interpreted as a quantitative representation of 
the dispersion problem discussed in \cite{Sha06}.

Eq.(\ref{eq:ciout}) shows how the two photon amplitudes 
$C_{11}$, $C_{12}$, and $C_r$ can be determined from the 
semi-classical description of the non-linearity in terms of $b^{(1)}(t)$ 
and $b^{(3)}(t)$. Specifically, the 
amplitude of the $\psi_1(t)$-component in $b^{(3)}(t)$ is equal to 
$C_{11}-1$ and the amplitude of the $\psi_2(t)$-component is equal to 
$C_{12}/\sqrt{2}$. Since $\psi_1(t)$ is equal to the linear component 
$b^{(1)}(t)$ of the semi-classical output, the $\psi_1(t)$-amplitude 
can be determined from the overlap integral of $b^{(3)}(t)$ and 
$b^{(1)}(t)$. The two photon amplitude $C_{11}$ is therefore 
given by
\begin{equation}
\label{eq:c11}
C_{11}=1+\int b^{(1)}(t)^* b^{(3)}(t)dt.
\end{equation}
As mentioned above, this parameter describes non-linear phase shifts that 
do not change the pulse shape of the photon wavepackets. This kind of 
single mode phase shift has been the focus of most of the previous work
on non-linear optical quantum gates. In fact, $C_{11}$ is equivalent to the 
parameter previously introduced by Koshino and Ishihara to evaluate the
performance of a quantum non-linearity based on a semi-classical result
\cite{Kos04}. Our analysis completes this approach by taking into account the
details of the pulse shape $b^{(3)}(t)$ that describes the semi-classical
effects of the third-order non-linearity.
In terms of the quantum mechanical description, this results
in the introduction of the additional output mode $\psi_2(t)$ and the
associated two photon amplitude $C_{12}$ describing the transfer 
of one photon to this new mode. The magnitude of this amplitude can be 
determined by taking the total intensity of $b^{(3)}(t)$ and subtracting 
the intensity accounted for by $C_{11}$,
\begin{eqnarray}
\label{eq:c12}
|C_{12}|^2&=& 2\left(\int |b^{(3)}(t)|^2 dt-|C_{11}-1|^2\right)
\nonumber \\
&=& 2\left(\int |b^{(3)}(t)|^2dt-\left|\int b^{(1)}(t)^*b^{(3)}(t)dt\right|^2\right).
\end{eqnarray}
The phase of $C_{12}$ depends on the definition of $\psi_2(t)$. It can therefore always be set to zero. 
Since the wavefunction $\psi_2(t)$ of the target mode can also be obtained from the pulse shape
$ b^{(3)}(t)$, the amplitude $C_{12}$ also describes a fully coherent process that may be used to
implement well controlled non-linear operations on optical quantum states.

All effects that cannot be described by the two photon amplitudes
$C_{11}$ and $C_{12}$ are summarized by the amplitude $C_r$. 
Since the two photon output wavefunction given by eq.(\ref{eq:2t2}) is 
normalized, $C_r$ can be obtained from
\begin{equation}
\label{eq:cr}
|C_{r}|^2=1-|C_{11}|^2-|C_{12}|^2
\end{equation}
Because this amplitude is associated with a two photon wavefunction
that cannot be described within the two mode expansion of section 
\ref{sec:system}, it has to be regarded as a source of decoherence in
any straightforward implementation of a non-linear quantum gate.
It thus provides a quantitative measure of the dispersion problems
raised in \cite{Sha06}, and it is an interesting question to what extend
the amplitude $C_r$ can be minimized while retaining the desired
non-linear effects described by $C_{11}$ and $C_{12}$.

%%---Quantum limit of semi-classical non-linearity
%%
%%%
\begin{figure}[th]
\begin{picture}(300,240)
\put(0,0){\makebox(300,240)
{
\scalebox{0.8}[0.8]{\includegraphics{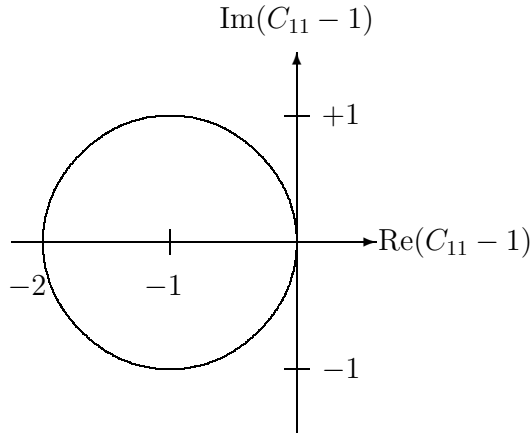}}}}
\end{picture}
\vspace*{-1.5cm}
\caption{\label{limit} Illustration of the quantum limit of the
integral $\int b^{(1)}(t)^* b^{(3)}(t)dt=C_{11}-1$ describing the overlap between the 
linear and the non-linear part of the semi-classical output pulse. 
}
\end{figure}
%%%
Before moving on to specific examples of quantum level non-linearities,
it may be interesting to consider the natural limits imposed on 
semi-classical third-order non-linearities by the amplitudes $C_{11}$ and
$C_{12}$ of the two photon output component. While classical physics imposes
no such limits, quantization makes it impossible to have non-linear effects 
for light field intensities far below the single photon level. In the 
present theory, this limit is expressed quantitatively as restrictions on
the magnitude of $b^{(3)}(t)$ corresponding to the requirement that
$|C_{11}|^2+|C_{12}|^2 \leq 1$. Considering only $C_{11}$, the overlap between
$b^{(3)}(t)$ and the normalized linear output $b^{(1)}(t)$ is limited by
\begin{equation}
\label{eq:limit1}
\left| \int b^{(1)}(t)^* b^{(3)}(t)dt + 1 \right| \leq 1.
\end{equation}
Figure \ref{limit} shows this limit of the overlap integral between the
non-linear and the linear output wavefunctions as a circle in the complex
plane. Note that the real part of the overlap is always negative, indicating 
that all third-order non-linearities reduce the output amplitude. 
Moreover, phase shifts of $|\phi|\leq \pi/2$ are associated with a 
minimal amplitude reduction of $(1-\cos\phi)$. 
Likewise, a change in pulse shape described by the photon interaction 
amplitude $C_{12}$ imposes a minimum on the negative value of the overlap
between $b^{(1)}(t)$ and $b^{(3)}(t)$. As $C_{12}$ increases, the radius of 
the circle in figure \ref{limit} is reduced to $\sqrt{1-|C_{12}|^2}$,
resulting in a minimal amplitude reduction of
\begin{equation}
\label{eq:limit2}
\mbox{Re} 
\left( \int b^{(1)}(t)^* b^{(3)}(t)dt \right) \leq 
\; -\left(1-\sqrt{1-|C_{12}|^2}\right).
\end{equation}
Thus, any non-linearity includes a reduction of the coherent output 
amplitude in the linear mode $\psi_1(t)= b^{(1)}(t)$. In particular, the
maximal non-linear change in pulse shape ($C_{12}=1$) requires an 
overlap of $-1$ between $b^{(1)}(t)$ and $b^{(3)}(t)$.

\section{Dynamics of the non-linear system}
\label{sec:dynamics}

As shown in the previous section, the two photon amplitudes $C_i$ 
can be determined from the output pulse shapes 
$b^{(1)}(t)$ and $b^{(3)}(t)$ obtained from a semi-classical analysis
of the non-linear optical response. It is therefore possible to determine
a significant part of the transformation acting on a two photon
wavepacket by solving the non-linear dynamics of the system in response
to a specific input pulse shape. 
In general, the dynamics of absorption and emission in an optical
system excited by a coherent light field pulse can be represented by 
the dynamics of the density matrix, where the excitation is driven by
a term linear in the coherent field amplitude $\alpha$. Initially,
the system is in its ground state $\rho^{(0)}$. Weak excitations are
described by a linear response $\rho^{(1)}$, such that the density matrix
dynamics are approximately given by $\rho(t)=\rho^{(0)}+\alpha \rho^{(1)}$.
The field $b_{\mbox{\small out}}(t)=\alpha b^{(1)}(t)$ emitted by 
the system is then determined by the dipole or field expectation values 
of $\rho^{(1)}$ according to input-output theory \cite{Wal}. 
To describe the non-linear response, the density matrix dynamics can be
expanded to include higher orders of $\alpha$. Specifically, the 
excitation of the density matrix caused by an input pulse of intensity
$|\alpha|^2$ can be described by a term $|\alpha|^2 \rho^{(2)}$.
However, the dipole or field expectations of this term are zero in all
symmetric systems, since this contribution to $\rho$ is invariant under
a change of sign in $\alpha$. Hence the lowest order non-linearity 
is obtained from the third-order term $\alpha |\alpha|^2 \rho^{(3)}$
that describes the effects of the excitations on the response of the
system. As the example given in the following will show, this kind of 
expansion can be done sequentially, resulting in a fairly simple
and straightforward integration of a series of linear response equations.

\begin{figure}[bth] 
\begin{center}
\begin{picture}(300,240)
\put(0,0){\makebox(300,240)
{\scalebox{0.9}[0.9]{\includegraphics{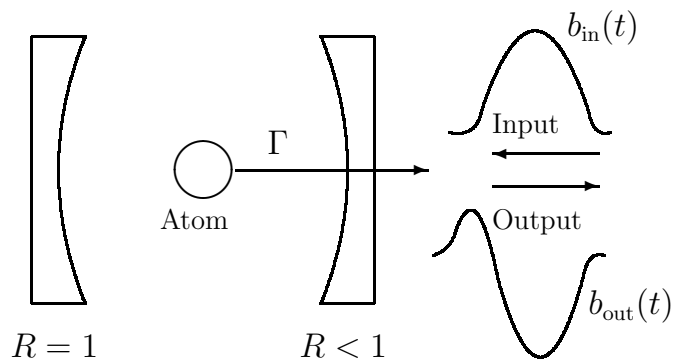}}}}
\end{picture}
\end{center}
\vspace{-2cm}
\caption{Schematic representation of an atom-cavity system. 
If the reflectivity $R$ of the back mirror is close to 1, the
cavity mediated emission determines the coupling $\Gamma$ with the
field reflected by the front mirror.
}
\label{cavity}
\end{figure}

One of the most important systems considered for the realization of
non-linear optical quantum gates is a single atom in a cavity.
As Turchette et al. demonstrated experimentally \cite{Tur95}, such a 
system exhibits a strong non-linear phase shift in the weak 
coupling regime, where the cavity dynamics can be adiabatically 
eliminated and the system response is described by the Bloch equations
of a two level atom. The problem of evaluating the strength
of the non-linear response in this system has recently attracted a lot of interest, 
largely motivated by the improved experimental 
possibilities due to advances in cavity design \cite{Aok06,Cha07,Tak08,Fus08,Min09,Wak06,Auf07,Xia08}.
Our theory permits us to analyze the temporal and spectral properties of this non-linear 
response in terms of realistic pulse shapes without the complications introduced by
a full quantum analysis of the entangled two photon wavefunction. It should be noted that we achieve 
this without any approximations, simply by omitting the parts of the two photon wavefunction 
that do not contribute to the average field $b_{\mbox{\small out}}(t)$. In fact, the third 
order solutions of the semi-classical Bloch equations discussed in the following can also be 
obtained from the two photon wavefunctions determined in \cite{Koj03} if one of the two photons 
is projected onto the single photon wavefunction. Thus the fully quantum mechanical analysis of the 
two photon response gives the same results for $b_{\mbox{\small out}}(t)$ as the semi-classical analysis. It only provides additional information about the elusive component 
$\mid \mbox{rest} \rangle$, which usually reqires an infinite number of additional modes for 
its precise characteization due to the spectral and temporal entanglement of the two photon 
wavefunction \cite{Hof03b}. For the coefficient $C_{11}$, the exact correspondence of the two 
photon response with the semi-classical result was first pointed out by Koshino and Ishihara in 
\cite{Kos04}, where they suggested the application of a coherent amplitude of $\alpha=1/\sqrt{2}$ 
to obtain the maximal semi-classical response corresponding to $C_{11}$. In addition to this 
evaluation of $C_{11}$, the more detailed analysis developed here allows us to
identify the photon transfer amplitude $C_{12}$ that describes the non-linear change of pulse shape
in terms of a conditional transition between the modes in the two photon output. 
As will be shown below, this effect is 
actually much stronger than the anticipated phase shift and may therefore
play a significant role in possible realizations of non-linear optical
quantum gates. Finally, we can also determine the precise pulse shapes
$\psi_1(t)$ and $\psi_2(t)$ that characterize the optical modes involved 
in the output of the quantum operation.

Fig. \ref{cavity} shows a schematic illustration of a single atom cavity system
and its input/output characteristics. An input pulse 
$b_{\mbox{\small in}}(t)$ is incident on the front mirror of the cavity.
The cavity field adiabatically couples the input pulse to the atom 
with a coupling strength corresponding to a cavity enhanced spontaneous 
emission rate of $2 \Gamma$. The atom is excited and re-emits the 
absorbed energy by dipole emission through the cavity.
Interference between this dipole emission and the reflected input 
pulse then results in the output pulse $b_{\mbox{\small out}}(t)$. 
In the following, we assume that the losses of the system are negligible,
so that all of the photons emitted by the atom will be found in
$b_{\mbox{\small out}}(t)$. 
The dipole response of the atom can be described by the well known
optical Bloch equations \cite{Hof03a,Wal},
\begin{eqnarray}
\label{eq:bloch1}
\frac{d}{dt} \langle \hat{\sigma}_{-} \rangle(t)&=&-\Gamma \langle \hat{\sigma}_{-} \rangle(t)-i2\sqrt{2\Gamma}\; \alpha\, b_{\mbox{\small in}}(t)\; \langle 
\hat{\sigma}_{z} \rangle(t) \\
\label{eq:bloch2}
\frac{d}{dt} \langle \hat{\sigma}_{z} \rangle(t) &=& -2\Gamma\left(
\langle \hat{\sigma}_{z} \rangle(t)+\frac{1}{2}\right) \nonumber \\ &&
 + i\sqrt{2\Gamma} \left(\alpha \,
 b_{\mbox{\small in}}(t)\langle \hat{\sigma}_{-} \rangle^*(t)
- \alpha^* \, b_{\mbox{\small in}}^{*}(t)\langle \hat{\sigma}_{-} \rangle(t)\right),
\end{eqnarray}
where $\hat{\sigma}_{-}$ is the operator describing the atomic dipole and
$\hat{\sigma}_{z}$ is the operator describing the excitation of the atom.
In general, the output field can be determined from the input field and
the corresponding dipole response $\langle \hat{\sigma}_{-} \rangle$ of
the atom according to input-output theory \cite{Hof03a,Wal}. For a lossless
system, the corresponding relation reads 
\begin{equation}
\label{eq:iotout}
b_{\mbox{\small out}}(t)=\alpha \, b_{\mbox{\small in}}(t)+i\sqrt{2\Gamma} 
\langle \sigma_{-}\rangle(t).
\end{equation}
In general, the dipole response $\langle \sigma_{-}\rangle(t)$ is a non-linear
function of the input amplitude $\alpha$. In order to determine the linear
and third-order response in $\alpha$, we can now apply the
procedure described at the beginning of this section to the Bloch
equations (\ref{eq:bloch1}) and (\ref{eq:bloch2}). 

Initially, the
atom is in the ground state, so the zero order density matrix is described
by $\langle \hat{\sigma}_{z}\rangle^{(0)}(t)= -\frac{1}{2}$. The linear response of the atom then determines the first order dipole term 
$\alpha \langle \hat{\sigma}_{-} \rangle^{(1)}$ according to the linear 
relaxation dynamics given by using 
$\langle \hat{\sigma}_{z}\rangle^{(0)}(t)= -\frac{1}{2}$ in 
eq.(\ref{eq:bloch1}),
\begin{equation}
\label{eq:first}
\frac{d}{dt} \langle \hat{\sigma}_{-} \rangle^{(1)}(t)=-\Gamma \langle 
\hat{\sigma}_{-} \rangle^{(1)}(t)+i\sqrt{2\Gamma}b_{\mbox{\small in}}(t).
\end{equation}
This equation can be solved for any input pulse shape by simply
integrating the linear response. In principle, the second order of the
density matrix is obtained by using the first order result
$\langle \hat{\sigma}_{-} \rangle^{(1)}$ as part of the excitation term in the 
relaxation dynamics described by eq.(\ref{eq:bloch2}). However, a comparison
of eq.(\ref{eq:bloch1}) and eq.(\ref{eq:bloch2}) shows that this equation
is always solved by the absolute square of the first
order dipole term, $\langle \hat{\sigma}_{z}^{(2)} \rangle = 
|\langle \hat{\sigma}_{-} \rangle^{(1)}|^2$.
We can then determine $\langle \hat{\sigma}_{-} \rangle^{(3)}(t)$ from 
eq.\ref{eq:bloch2} by using $\langle \hat{\sigma}_{z} \rangle^{(2)}(t)$
instead of $\langle \hat{\sigma}_{z} \rangle^{(0)}(t)$,
\begin{equation}
\label{eq:third}
\frac{d}{dt}\langle \hat{\sigma}_{-} \rangle^{(3)}(t) =-\Gamma \langle 
\hat{\sigma}_{-} \rangle^{(3)}(t)-i2\sqrt{2\Gamma}\; b_{\mbox{\small in}}(t) 
\; |\langle\hat{\sigma}_{-}\rangle^{(1)}(t)|^2.
\end{equation}
This equation has the same form as eq.(\ref{eq:first}) and can therefore be
solved by the same kind of integration, corresponding to the linear response 
of the dipole to a modified input pulse of 
$-2 b_{\mbox{\small in}}|\langle \hat{\sigma}_{-}\rangle^{(1)}|^2$.
Specifically, the third-order response describes the reduction of the linear
response by the saturation of the gradually excited atoms, as indicated by the
negative sign of the modified input pulse.

Having obtained the linear and third-order responses of the atomic dipole, we
can now express the output field up to third order in $\alpha$ using input-output
theory as given by eq.(\ref{eq:iotout}). The result reads
\begin{equation}
\label{eq:1atomout}
b_{\mbox{\small out}}(t)= \alpha \left( b_{\mbox{\small in}}(t)
+i\sqrt{2\Gamma} \langle \hat{\sigma}_{-} \rangle^{(1)}(t)\right)
+\alpha |\alpha|^2 \left( i \sqrt{2\Gamma} \langle \hat{\sigma}_{-}\rangle^{(3)}(t)\right)
\end{equation}
Comparison with eq.(\ref{eq:nlout}) shows how the solutions for the expectation
values of the atomic dipole define the linear and non-linear output pulse
shapes $b^{(1)}(t)$ and $b^{(3)}(t)$. We can thus obtain the semi-classical
characterization of the optical non-linearity necessary for the determination of
the two photon amplitudes $C_i$ and the wavefunctions $\psi_1(t)$ and $\psi_2(t)$
describing the modes used for the quantization by solving the density matrix
dynamics of the system up to third-order in the coherent excitation amplitude
$\alpha$.

\section{Pulse duration dependence of two photon amplitudes}
\label{sec:tdependence}

In general, the two photon amplitudes $C_i$ depend on the specific shape of the
input pulse defined by $b_{\mbox{\small in}}(t)$. In the following, we focus on 
resonant pulses, since such pulses seem to be the most promising candidates for 
strong non-linear effects \cite{Hof03a}. For a fixed pulse shape, the non-linearity then depends
only on the ratio between pulse duration $T$ and the relaxation time $1/\Gamma$
of the atomic dipole. By applying our method of analysis, we can determine this dependence
of non-linear effects on the scaled pulse duration $\Gamma T$. 

To cover a sufficiently wide range of possible pulse shapes while keeping the 
calculations relatively simple and efficient, we have chosen the four pulse
shapes given in table \ref{table1}. The most notable difference between the 
pulse shapes is that the change of the field is not continuous for the rectangular
and the rising exponential pulse, while it is continuous for the symmetric
exponential pulse and the Gaussian pulse. Moreover, only the rising exponential 
pulse is not symmetric around its peak. It should thus be possible to get insights
into the effects of discontinuities and symmetry on the non-linear transformation
of the pulses. 

\begin{table}[th]
\caption{\label{table1}
Definition of input pulse shapes $b_{\mbox{\small in}}(t)$.
}
\begin{tabular}{|clc|crclc|}
\hline \hline
&&&&&&
\\[-0.4cm]
& Type of pulse &&&
\multicolumn{3}{c}{Wavefunction for pulse duration $T$} &
\\[0.1cm] \hline \hline
&&&&&&
\\[-0.4cm]
& Rectangular pulse &&&
$b_{\mbox{\small in}}(t)$ &=& 
$\left\{
\begin{array}{cl}
1/\sqrt{T} &\mbox{for}\hspace{0.2cm} -T<t<0
\\
0 &\mbox{else}
\end{array}
\right.$ &
\\[0.2cm] \hline
&&&&&&&
\\[-0.2cm]
& Rising exponential pulse &&&
$b_{\mbox{\small in}}(t)$ &=& 
$\left\{
\begin{array}{cl}
\sqrt{2/T} \exp(t/T) &\mbox{for}\hspace{0.2cm} t<0
\\
0 &\mbox{for} \hspace{0.2cm} t>0
\end{array}
\right.$ &
\\[0.2cm] \hline
&&&&&&&
\\[-0.2cm]
& Symmetric exponential pulse &&&
$b_{\mbox{\small in}}(t)$ &=& 
$\sqrt{2/T} \exp(-2|t|/T)$ &
\\[0.2cm] \hline
&&&&&&&
\\[-0.2cm]
& Gaussian pulse &&&
$b_{\mbox{\small in}}(t)$ &=& 
$\sqrt{2/(\sqrt{\pi} T)} \exp(-2 \; t^2/T^2)$ &
\\[-0.3cm]
&&&&&&&
\\ \hline
\end{tabular}
\end{table}

\begin{figure}[th]
\begin{picture}(490,300)
\put(0,0){\makebox(470,300)
{\scalebox{0.9}[0.9]{\includegraphics{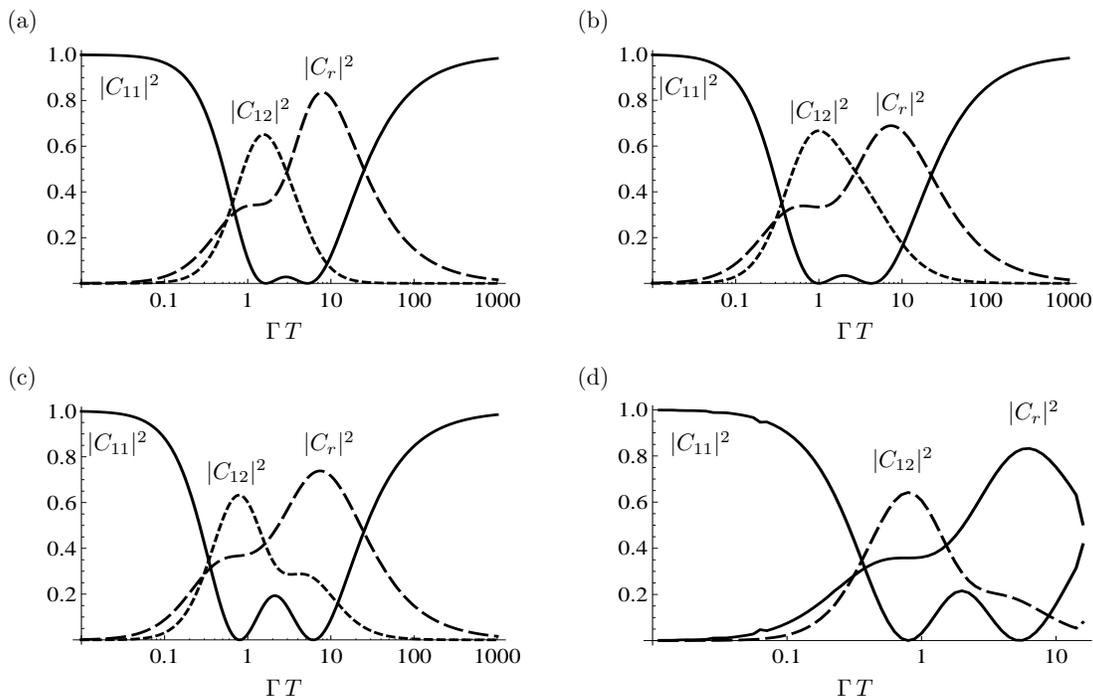}}}}
\end{picture}
\caption{\label{pulsetime}
Dependence of the squared two photon amplitudes $|C_i|^2$ on pulse duration $\Gamma T$ 
for (a) rectangular input pulses, (b) rising exponential input pulses,
(c) symmetric exponential input pulses, and (d) Gaussian input pulses. Note that the pulse
duration is given on a logarithmic scale.
}
\end{figure}

For any given pulse shape and pulse duration, the non-linear output wavefunctions can be determined by solving eqs.(\ref{eq:first}-\ref{eq:1atomout}). Using these results, it is then possible to determine the two photon amplitudes 
$C_{11}$, $C_{12}$, and $C_r$ from eqs. (\ref{eq:c11}-\ref{eq:cr}). 
The dependence of the results on pulse duration $T$ for each of the four input 
pulse shapes is shown in fig. \ref{pulsetime}.  
All four pulse shapes show a very similar pulse duration dependence, 
indicating that the two photon amplitudes $C_i$ do not depend much on the specific pulse shape. 
Significant non-linear effects can typically be observed for pulse durations
between about $0.1/\Gamma$ and $100/\Gamma$. For shorter pulses ($\Gamma T<0.1$), 
the bandwidth is too broad for resonant absorption, and for longer pulses
($\Gamma T>100$), the photon density is too low for efficient non-linear 
interactions. Between these limits, the amplitude $C_{11}$ drops from its linear
value of $1$ to negative values representing the non-linear phase flip originally
proposed for use in non-linear optical quantum gates \cite{Tur95,Hof03a,Kos04}.
However, even the maximal values of the negative amplitude $C_{11}$ represent only 
small fractions of the two photon output wavefunction. For the symmetric exponential
input pulse (fig. \ref{pulsetime}(c)) and the Gaussian input pulse (fig. 
\ref{pulsetime}(d)), the maximum is at about 0.2, while it is below 0.05 for the
discontinuous pulse shapes of the rectangular input pulse (fig. \ref{pulsetime}(a))
and the rising exponential pulse (fig. \ref{pulsetime}(b)). Thus, the non-linear
phase flip is both limited in magnitude, and sensitive to discontinuities in the
input pulse shape.

On the other hand, 
the squared amplitude $|C_{12}|^2$ describing the probability of a non-linear 
transfer of exactly one photon to an orthogonal output mode $\psi_2(t)$ has a peak
value of about $2/3$ for all four pulse shapes. This means that the conditional
photon transfer is more efficient and less sensitive to pulse shape effects such
as discontinuities. It may therefore be useful to consider this effect as an
alternative option for the realization of non-linear
optical quantum gates.

\section{Analysis of output pulse shapes at maximal photon transfer probability}
\label{sec:pulse}

Since the conditional photon transfer described by $C_{12}$ is the strongest
non-linear effect modifying the two photon output state, it may be useful to 
take a closer look at the coherent pulse shapes $\psi_1(t)$ and $\psi_2(t)$ that
determine about two thirds of the two photon output state. These pulse shapes can
be obtained from the semi-classical results for $b^{(1)}(t)$ and $b^{(3)}(t)$
using the amplitudes $C_{11}$ and $C_{12}$ and eqs. (\ref{eq:nlout}) and 
(\ref{eq:ciout}). The maximal values of $C_{12}$ are (a) $|C_{12}|^2=0.66$
at $\Gamma T=1.56$ for rectangular input pulses, (b) $|C_{12}|^2=2/3$
at $\Gamma T=1$ for rising exponential input pulses, (c) $|C_{12}|^2=0.67$
at $\Gamma T=0.78$ for symmetric input pulses, and (d) $|C_{12}|^2=0.64$
at $\Gamma T=2$ for Gaussian input pulses. For these pulse durations, the values 
of $C_{11}$ are all vanishingly small, so the probability of finding both photons in the linear 
output mode is close to zero. 
 
\begin{figure}[th]
\begin{picture}(490,300)
\put(0,0){\makebox(470,300)
{\scalebox{0.9}[0.9]{\includegraphics{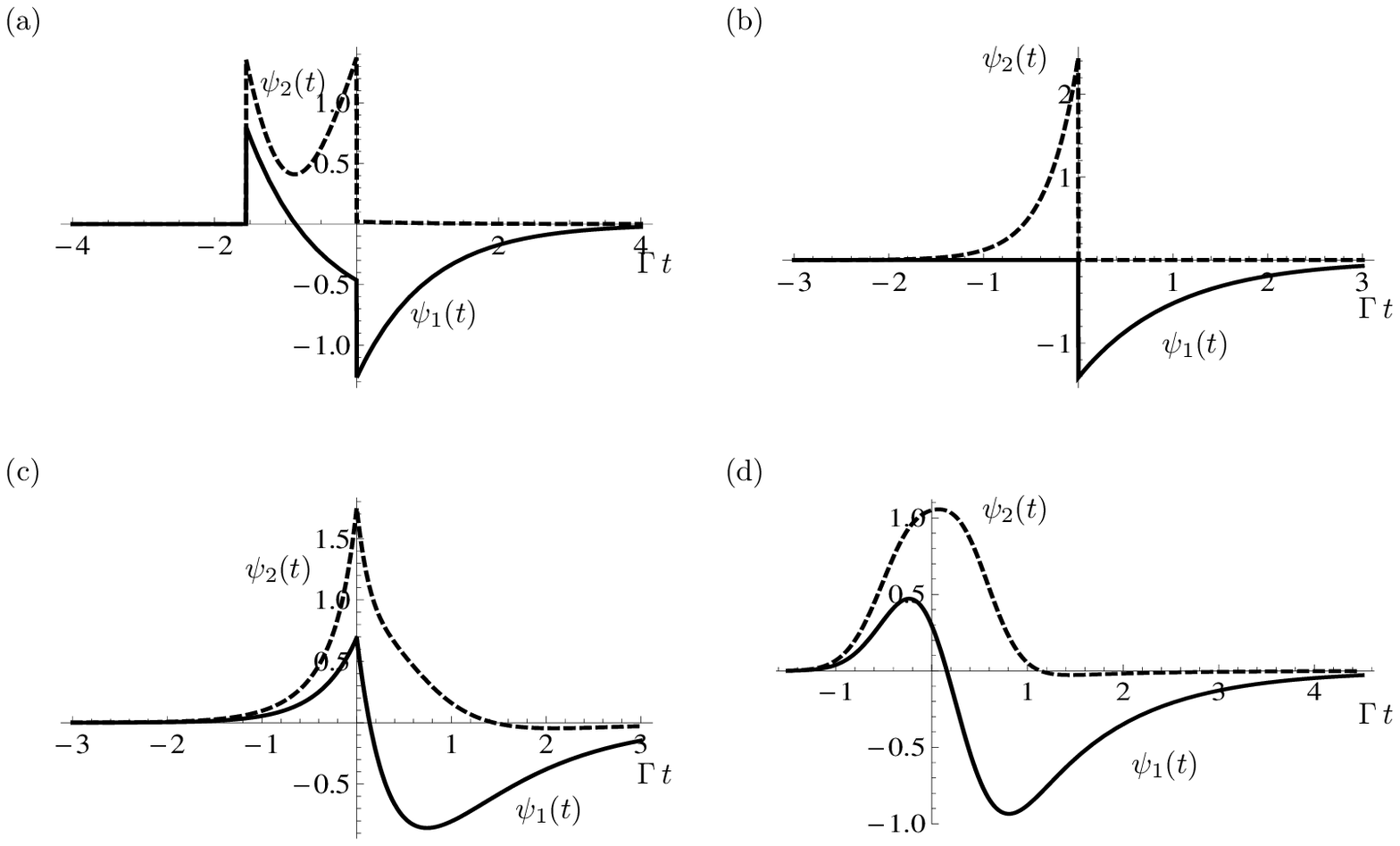}}}}
\end{picture}
\caption{\label{pulseshapes}
Output pulse shapes $\psi_1(t)$ and $\psi_2(t)$ describing the linear output wavefunction and the target pulse of the conditional single photon transfer
associated with $C_{12}$. (a) shows the pulse shapes for a rectangular 
input pulse, (b) for a rising exponential input pulse, 
(c) for a symmetric exponential input pulses, and (d) for a Gaussian input pulse.
}
\end{figure}

Fig. \ref{pulseshapes} shows the output pulse shapes at the maximal mode transfer
amplitudes given above. Interestingly, the shapes of the non-linear target modes
$\psi_2(t)$ are somewhat similar in shape to the original input modes, while the 
wavefunctions $\psi_1(t)$ are all delayed and change their sign to negative amplitudes
after an initial positive peak (except for the rising exponential input in 
fig. \ref{pulseshapes} (b), where the
amplitude is exactly zero until $t=0$). This observation suggests a simple intuitive 
explanation for the non-linear mode transfer effect described by $C_{12}$. 
If the pulse duration is perfectly matched to the
absorption time, the atom is excited by one of the two photons for the entire pulse
duration. Therefore, the second photon cannot be absorbed and passes the atom 
without the changes to its wavefunction otherwise induced by the linear dipole 
dynamics. If the transmitted mode $\psi_2(t)$ is orthogonal to the linear response
mode $\psi_1(t)$, then the two photon output wavefunction will have exactly one photon
in each of the two modes.  
A particularly interesting case may be that of the rising exponential input pulse
shown in fig. \ref{pulseshapes} (b). Here, the input pulse and the output pulse are
clearly separated in time. It may therefore be possible to distinguish the two modes
by a sufficiently fast time-dependent gate. Specifically, any photon detected before
$t=0$ indicates that a second photon was absorbed by the atom, since the linear 
output is exactly zero for $t<0$. This effect could be used for a highly efficient 
elimination of multi-photon components in single mode quantum states.

\section{Conclusions}
\label{sec:conclusions}
We have presented an analysis of the relation between the average field amplitudes
obtained from the density matrix dynamics of coherently driven systems and
the transformation of two photon quantum states by a non-linear system. The results show
how the spectral and temporal features of non-linear field transformations affect
the performances of non-linear quantum gates operating on few photon states. 
It is thus possible to predict whether a given non-linear system can be used to 
implement a non-linear optical quantum gate operating on 
superpositions of quantum states with zero, one, and two photons, based on quantitative
data of the intensity dependence of the coherent output field. In general, such data can
be determined either experimentally, using coherent light inputs and homodyne detection, or
theoretically, by solving the density matrix dynamics.

The application to a resonant single-atom non-linearity shows that the most promising
non-linearity may not be the widely investigated two photon phase shift, but a non-linear
photon transfer process to a two photon output wavefunction where exactly one of
the two photons is in a mode orthogonal to the single photon output mode. 
The investigation of the specific pulse shape shows that the effect may be understood
in terms of the saturation of the atom when the linear output wavefunction 
$\psi_1(t)$ is approximately orthogonal to the transmitted wavefunction $\psi_2(t)$.
If the two wavefunctions can be separated by appropriate time-dependent gates,
it may be possible to implement optical quantum gates based on this fundamental property of
single atom non-linearities.

\section*{Acknowledgements}
We would like to thank Mr. S. Ohnishi for his help with the numerical calculations. 
Part of this work has been supported by the Grant-in-Aid program of the Japanese
Society for the Promotion of Science, JSPS.


\begin{thebibliography}{xyz00}

%%--quantum non-linearities

\bibitem{Imo85}
N. Imoto, H. A. Haus, and Y. Yamamoto, Phys. Rev. A {\bf 32}, 2287 (1985).

\bibitem{Tur95}
Q. A. Turchette, C. J. Hood, W. Lange, H. Mabuchi, and H. J. Kimble, Phys. Rev. Lett. {\bf 75}, 4710 (1995).

\bibitem{Aok06}
T. Aoki, B. Dayan, E. Wilcut, W. P. Bowen, A. S. Parkins, T. J. Kippenberg, K. J. Vahala and H. J. Kimble, Nature {\bf 443}, 671 (2006).

\bibitem{Cha07}
D. E. Chang, A. S. Sorensen, E. A. Demler, M. D. Lukin,
Nature Physics {\bf 3}, 807 (2007). 

\bibitem{Tak08}
H. Takashima, H. Fujiwara, S. Takeuchi, K. Sasaki, and M. Takahashi,
Appl. Phys. Lett. {\bf 92}, 071115 (2008).

\bibitem{Fus08}
I. Fushman, D. Englund, A. Faraon, N. Stoltz, P. Petroff, and J. Vuckovic,
Science {\bf 320}, 769 (2008).

\bibitem{Min09}
B. K. Min, E. Ostby, V. Sorger, E. Ulin-Avila, L. Yang,
X. Zhang, and K. Vahala, Nature {\bf 457}, 455 (2009).

%%--spatiotemporal

\bibitem{Hof03a}
H. F. Hofmann, K. Kojima, S. Takeuchi, and K. Sasaki, J. Opt. B: Quantum Semiclass. Opt. {\bf 5}, 218 (2003). 

\bibitem{Koj03}
K. Kojima, H. F. Hofmann, S. Takeuchi, and K. Sasaki, 
Phys.Rev.A {\bf 68}, 013803 (2003).

\bibitem{Hof03b}
H. F. Hofmann, K. Kojima, S. Takeuchi, and K. Sasaki, 
Phys.Rev.A {\bf 68}, 043813 (2003).

\bibitem{Kos04}
K. Koshino and H. Ishihara, Phys. Rev. Lett. {\bf 93}, 173601 (2004).

\bibitem{Wak06}
E. Waks and J. Vuckovic, Phys. Rev. A {\bf 73}, 041803(R) (2006).

\bibitem{Auf07}
A. Auffeves-Garnier, C. Simon, J.-M. Gerard, and J.-P. Poizat,
Phys. Rev. A {\bf 75}, 053823 (2007). 

\bibitem{Kos07}
K. Koshino, Phys. Rev. Lett. {\bf 98}, 223902 (2007).

\bibitem{Koj07}
K. Kojima and A. Tomita, Phys. Rev. A 75, 032320 (2007)

\bibitem{Kos08}
K. Koshino, Phys. Rev. A {\bf 78}, 023820 (2008).

\bibitem{Xia08}
Y.-F. Xiao, S. K. \"Ozdemir, V. Gaddam, C.-H. Dong, N. Imoto, and L. Yang,
Opt. Express {\bf 16}, 21462 (2008).

\bibitem{Sha06}
J. H. Shapiro, Phys. Rev. A 73, 062305 (2006).

\bibitem{Sha07}
J. H. Shapiro and M. Razavi, New J. Phys. {\bf 9}, 16 (2007).

\bibitem{Leu08}
P. M. Leung, T. C. Ralph, W. J. Munro, and K. Nemoto, e-print arXiv: 0810.2828.

\bibitem{Koj04}
K. Kojima, H. F. Hofmann, S. Takeuchi, and K. Sasaki,
Phys. Rev. A {\bf 70}, 013810 (2004).

%%--experiments

\bibitem{Che06}
Y.-F. Chen, C.-Y. Wang, S.-H. Wang, and I. A. Yu, Phys. Rev. Lett. {\bf 96},
043603 (2006).

\bibitem{Mat09}
N. Matsuda, R. Shimizu, Y. Mitsumori, H. Kosaka, and K. Edamatsu,
Nature Photonics {\bf 3}, 95 (2009).

%--Bloch eq.

\bibitem{Wal}
D. F. Walls and G. J. Milburn, {\it Quantum Optics} (Springer, Berlin, 1994). 

\end{thebibliography}
\end{document}